\documentclass[doublespacing]{elsart}

\usepackage{graphicx}
\newcommand{\mc}{\;\mbox{,}}

\begin{document}
\begin{frontmatter}
\title {On the  Superconductivity in the Induced Pairing Model}

\author{R. Micnas$^1$}{\corauthref{1}}
\author{S. Robaszkiewicz$^1$}
\author{and A. Bussmann-Holder$^2$}

\corauth[1]{Corresponding Author: 
         Roman Micnas 
	 Institute of Physics,
	 A. Mickiewicz  University,
         Umultowska 85, 
	 PL-61614 Poznan, Poland;
	 Phone: (+48 61) 8295-041, e-mail: rom@alpha.amu.edu.pl} 

\address{$^1$Institute of Physics,  A. Mickiewicz  University,
             Umultowska 85, 61-614 Pozna\'{n}, Poland }
\address{$^2$ Max-Planck Institut f\"ur Festk\"orperforschung, Heisenbergstrasse1, D-70569 Stuttgart,
Germany}

\begin{abstract} 
The two component model  
 of 
 coexisting 
local electron pairs and itinerant fermions  coupled 
via charge exchange mechanism, which mutually induces superconductivity in both
subsystems, is discussed.
The cases of isotropic $s$-wave and anisotropic pairing of extended $s$ and
$d_{x^2-y^2}$ -wave  symmetries are   analyzed
for a 2D square lattice  within the BCS-mean field approximation 
and the Kosterlitz-Thouless theory.
We determined the phase diagrams 
and superconducting characteristics as a function of 
the position of the local pair (LP) level and the total
electron concentration. The model exhibits several types of interesting 
crossovers from BCS like behavior
to that of  LP's. 
Some of our results are discussed in 
connection with  a two-component scenario 
of preformed pairs and unpaired electrons 
 for exotic
superconductors.
\end{abstract}


\end{frontmatter}

A system
of interacting charged bosons 
  (bound electron pairs)  and  electrons  can  show  features  which  are 
  intermediate between those of  local pair  superconductors and those 
  of classical BCS systems. Such a  two component (boson-fermion) model is of relevance
  for high temperature superconductors (HTS) and other exotic superconductors
\cite{{Micnas90},{gorkov87},{rob87},{lee89},{jrmr96},{NATO},{larkin97},{castro},{rmsrbt},{unpublished}}.
  Recently, we have studied a generalization of this model to the case of
  anisotropic pairing \cite{rmsrbt,unpublished}. Here, we shortly outline 
 the study 
 and  present further results concerning the phase diagrams
  and superconducting properties of such a system  in the case of isotropic $s$-wave pairing
  as well as for anisotropic pairings of $d$- and extended $s$-wave symmetries, 
  for a 2D square lattice.
 
 \section{The Model}
 We consider the model of coexisting
 electron pairs (formed by $"d"$ electrons) and itinerant $"c"$ electrons defined by 
 the following effective   Hamiltonian 
\begin{eqnarray}\label{main}
{\it H} =  \sum_{\bf {k}\sigma} (\epsilon_{\bf k}-
\mu)c^{\dagger}_{\bf{k}\sigma}
c_{\bf {k}\sigma} + 2\sum_{i}(\Delta_{0}-\mu)b^{\dagger}_{i}b_{i}
-\sum_{ij}J_{ij}b^{\dagger}_{i}b_{j}
+\nonumber \\
\sum_{\bf k, q}\left[V_{{\bf q}}({\bf k}) c^{\dagger}_{\bf{k} +\bf {q}/2,
\uparrow}c^{\dagger}_{{\bf {-k} +\bf {q}}/2, \downarrow} b_{\bf {q}} 
+ H.c.\right],
\end{eqnarray}
where $\epsilon_{\bf k}$  refers to  the   band energy of  the  c-electrons, 
  $\Delta_{0}$   measures  the 
  relative position of the LP level with respect to the bottom of 
  the c-electron band, 
  $\mu$ is the chemical potential which ensures that the 
  total number of particles in the system is constant, i.e. 
$ n=\frac{1}{N}(\sum_{\bf k\sigma}
<c^{\dagger}_{\bf k \sigma}c_{\bf k \sigma}>
+2\sum_{i} <b^{\dagger}_{i}b_{i}>)=n_{c}+2n_{B}.$
$n_{c}$ is the concentration of $c$-electrons, $n_{B}$
is the number of
$d$-pairs (hard-core bosons) per site. $J_{ij}$ is the pair hopping integral 
The 
operators for local pairs 
$ {b^{\dagger}_{i},b_{i}}$ 
obey the Pauli spin 1/2 commutation rules. $V_{\bf q}(\bf k)$ describes the coupling
between the two subsystems. Furthermore,we set $J_{ij}=0$. We  will analyze the case 
$V_{{\bf q}}({\bf k})=V_{0}({\bf k})= I \phi_{k}/\sqrt{N}$, 
and neglect its $q$ dependence  at small $q$. 
The interaction term
takes  the form of coupling,   
via the center of mass momenta ${\bf q}$,  of the singlet pair of $c$-electrons
$B^{\dagger}_{\bf q}$
 and the hard-core boson $b_{\bf q}$:
 \begin{eqnarray}
H_{1}= \frac{1}{\sqrt{N}}\sum_{\bf q}I(B^{\dagger}_{\bf q}b_{\bf q} 
+b^{\dagger}_{\bf q}B_{\bf q}).
\end{eqnarray}
$B^{\dagger}_{\bf q}=\sum_{\bf k}\phi_{k}c^{\dagger}_{\bf{k} +\bf {q}/2,
 \uparrow}c^{\dagger}_{\bf {-k} +\bf {q}/2, \downarrow}$ denotes 
 the singlet pair creation
operator of $c$-electrons and $I$ is the coupling constant.
 The pairing symmetry, on a 2D square lattice,  is determined by the form of 
 $\phi_{k}$, 
 which is $1$  for  on-site pairing ($s$),
 $\phi_{k}= \gamma_{k}=\cos(k_{x}a)+\cos(k_{y}a)$ for  extended $s$-wave
 ($s^{*}$) and $\phi_{k}= \eta_{k}=\cos(k_{x}a)-\cos(k_{y}a)$ for  
 $d_{x^2-y^2}$-wave pairing ($d$).
In general, one can consider a decomposition 
 $I\phi_{k}=g_{0}+g_{s}\gamma_{k}+g_{d}\eta_{k}$, 
with appropriate coupling parameters for different 
 symmetry channels.

The numerical results presented below are based on  the BCS-Mean-Field Approximation (MFA) and the
 Kosterlitz-Thouless (KT) theory for 2D superfluid. 
In the BCS-MFA  the free energy of the system is calculated to be:
\begin{eqnarray}\label{freeenergy}
F/N = -\frac{2}{\beta N}\sum_{{\bf k}}\ln\left[2\cosh(\beta E_{{\bf
k}}/2)\right]
-\frac{1}{\beta}\ln\left[2\cosh(\beta\Delta)\right] + C ,\\
C = -\epsilon_{b} +\Delta_{0}+\mu (n_{c}+2n_{B})-2\mu -2I |x_{0}|\rho_{0}^{x}~,
\end{eqnarray}
where the quasiparticle energy of the $c$-electron subsystem 
is given by \\$E_{\bf k}=
\sqrt{\bar\epsilon_{\bf k}^{2}+\bar{\Delta}_{\bf k}^{2}}$, and $\bar\epsilon_{\bf k}=\epsilon_{\bf k}-\mu$, 
$\bar{\Delta}_{\bf k}^2=I^2\phi_{k}^2(\rho_{0}^{x})^2$, 
$\Delta=\sqrt{(\Delta_{0}-\mu)^2+I^2|x_{0}|^2}$,  $\beta=1/k_{B}T.$ 
The $c$-electron  dispersion is  
$\epsilon_{\bf k}~=~\tilde{\epsilon}_{\bf k}-\epsilon_{b}=
-2t\left[\cos (k_{x}a)+\cos (k_{y}a)\right]-
4t_{2}\cos (k_{x}a)\cos (k_{y}a)-\epsilon_{b}$,
with the  nn and nnn hopping parameters 
$t$ and $t_{2}$, respectively, $\epsilon_{b}=min \tilde{\epsilon}_{\bf k}$ .
It should be noted that the energy gap in the $c$-band  is due 
to nonzero Bose condensate amplitude ($|<b>|\neq 0$),  
and well defined Bogolyubov quasiparticles exist in the superconducting phase.
 The energy spectrum of the system is characterized by $E_{\bf k}$
and $\Delta$.
The superconducting  order parameters: $x_{0}=
\frac{1}{N}\sum_{k}\phi_{k}<c^{\dagger}_{k\uparrow}c^{\dagger}_{-k\downarrow}>$
 and $\rho^{x}_{0}=\frac{1}{2N}\sum_{i}<b^{\dagger}_{i}+b_{i}>$ and the chemical
 potential  $\mu$ are given by
 \begin{eqnarray}\label{orderpar}
 \frac{\partial F}{\partial x_{0}}=0,~~ 
 \frac{\partial F}{\partial \rho_{0}^{x}}=0,~~
  \frac{\partial F}{\partial \mu}=0.
 \end{eqnarray}
The superfluid density  derived within the linear response method and BCS theory  
is of the form:
\begin{eqnarray}\label{stiffness}
\rho_{s}=\frac{1}{2N}\sum_{k}
\left\{
\left(\frac{\partial\epsilon_{\bf k}}{\partial k_{x}}\right)^{2}
\frac{\partial f(E_{\bf k})}{\partial E_{\bf k}}
+\frac{1}{2}\frac{\partial^{2}\epsilon_{\bf k}}{\partial k_{x}^{2}}
\left[
1-\frac{\bar\epsilon_{\bf k}}{E_{\bf k}}
\tanh\left(\frac{\beta E_{\bf k}}{2}\right)
\right]
\right\} \mc
\end{eqnarray}
where  $f(E_{k})=1/\left[\exp(\beta E_{k})+1\right]$ is the Fermi-Dirac
distribution function. 
In the local limit: $\lambda^{-2}=(16\pi e^2/\hbar^2c^2)\rho_{s}$, where 
$\lambda$ is the London penetration depth.\\
Let us also consider the pair propagator for $c$-electron subsystem
\begin{eqnarray}
G_{2}({\bf q},\omega)=\frac{1}{N}<<B_{{\bf q}}|B^{\dagger}_{\bf q}>>_{\omega},
\end{eqnarray} 
where 
$<<B_{{\bf q}}|B^{\dagger}_{\bf q}>>_{\omega}$
is the time Fourier transform of the Green's function:
$ -i\Theta(t-t')<[B_{\bf q}(t),B^{\dagger}_{\bf q}(t')]>$.
In the normal state, using the equation of motion technique and 
 Random Phase Approximation one gets:
\begin{eqnarray}
G_{2}({\bf q},\omega)=\frac{\chi({\bf q}, \omega)}
{1-v_{eff}(\omega)\chi({\bf q}, \omega)}~,
\end{eqnarray}
where
\begin{eqnarray}
\chi({\bf q}, \omega)=\frac{1}{N}\sum_{\bf k}\phi^2_{k}
\frac{1-f(\bar\epsilon_{k+q/2})-f(\bar\epsilon_{k-q/2})}
{\omega-\bar\epsilon_{k+q/2}-\bar\epsilon_{k-q/2}}~,\\
v_{eff}(\omega)=\frac{I^2}{\omega-2(\Delta_{0}-\mu)}\tanh\left[
\beta(\Delta_{0}-\mu)\right]~,\label{effint}
\end{eqnarray}
$\chi({\bf q}, \omega)$ is the pair susceptibility and 
$v_{eff}(\omega)$-the effective interaction mediated by LP's.
Introducing the generalized T-matrix $\Gamma({\bf q}, \omega)$ via the relation
$G_{2}=\chi\Gamma/v_{eff}$,  one has
\begin{eqnarray}
\Gamma({\bf q}, \omega)=\frac{v_{eff}(\omega)}{1-v_{eff}(\omega)\chi({\bf q},
\omega)}~.
\end{eqnarray}
An instability of the normal phase occcurs at $\Gamma^{-1}(0,0)=0$ 
(the Thouless criterion),    
and $T_{c}$ is given by
\begin{eqnarray}\label{tcmfa}
1=\frac{I^2}{2(\Delta_{0}-\mu)}\tanh\left[\beta_{c}^{MFA}(\Delta_{0}-\mu)\right]
\frac{1}{N}\sum_{\bf k}
\phi^2_{k}\frac{\tanh\left(\beta_{c}^{MFA}\bar\epsilon_{\bf k}/2\right)}
{2\bar\epsilon_{\bf k}},
\end{eqnarray}
in full
agreement with the MFA  expression which one obtains from Eqs.({\ref{orderpar}), if 
$x_{0}\rightarrow 0,\rho_{0}^{x}\rightarrow 0$.\\

The MFA  transition
temperature ($T_{c}^{MFA}$) at which the gap amplitude vanishes
yields an estimation
of the $c$-electron pair formation temperature \cite{Micnas90}.
Due to the fluctuation effects the superconducting
phase transition will occur at a critical temperature  lower
than that given by the BCS-MFA theory.  In 2D, $T_{c}$ can be derived 
within the
KT theory for superfluid \cite{KT}, which describes the transition 
in terms of vortex-antivortex
pair unbinding.    We evaluate $T_{c}$
 using the KT relation for the universal jump of
the superfluid density $\rho_{s}$ at $T_{c}$ \cite{KT,Den} :
\begin{equation}
\frac{2}{\pi}k_{B}T_{c}=\rho_{s}(T_{c}),
\end{equation}
where $\rho_{s}(T)$ is given by Eq.(\ref{stiffness}) and  $x_{0}(T), 
\rho_{0}^{x}(T), \mu (T)$ are given by 
Eqs.(\ref{orderpar}). 
Hence, the critical temperature denoted further by $T_{c}^{KT}$ is determined from the
set of four self-consistent equations.
\section{Results and discussion}
We have performed a comprehensive 
analysis of the phase diagrams and 
superfluid properties of the model (\ref{main}) for different pairing
symmetries \cite{unpublished}. Below we  will discuss  
these results including the
supplementary ones.\\  
In Fig.1 we show the ground state diagrams as a function of $n$ and 
$\Delta_{0}/D$, plotted for $|I_{0}|=0$ ($I=-|I_{0}|$) and two fixed values of $t_{2}/t$.
A weak intersubsystem coupling $|I_{0}|$ will not change 
much characteristic lines 
of the diagrams, while increasing $|I_{0}|$ expands the range of the LP+E regime
\cite{czart01}.
Depending on the relative concentration of "c" and "d"  electrons 
  we distinguish three essentially different physical situations (in the absence
  of interactions). For $n\leq2$ it  will be:\\ 
  (i) $\Delta_{0} <0$ such that at $T=0K$ all the available electrons form 
local pairs (of   "d" 
  electrons) (the "local pair" regime, $2n_{B}\gg n_{c}$) (LP); \\
  (ii) $\Delta_{0} >0$ such that the "c" electron band 
  is filled  up  to  the 
  Fermi 
  level $\mu=\Delta_{0}$  and the remaining electrons are in the 
  form of  local pairs  (the  c+d regime or Mixed, $0<2n_{B},n_{c}<2)$
  (LP+E);\\ 
  (iii) $\Delta_{0} >0$ such that the Fermi level  $\mu<\Delta_{0}$ 
    and  consequently  at $T=0$K all  the 
  available electrons occupy the "c" electron states (the c-regime or"BCS",
$n_{c}\gg 2n_{B}$) (E).

For  $|I_{0}|\neq 0$,  in  the  case  (ii)  superconductivity  
is due to the  
  interchange between local pairs of  "d"  electrons  and  pairs  of  "c" 
  electrons. In this process "c" electrons become "polarized" into Cooper 
  pairs and "d" electron pairs increase their mobility by  decaying  into 
  "c" electron pairs. To this intermediate case neither the  standard  BCS 
  picture nor the picture of local pairs  applies  and  superconductivity 
  has a "mixed" character.
  The system shows features which are intermediate between the BCS and
preformed local pair regime. This regards the energy 
gap in the single-electron
excitation spectrum ($E_{g}(0))$, the $k_{B}T_{c}/E_{g}(0)$ ratio, the 
critical fields,
the Ginzburg ratio $\kappa$, the width of the critical regime 
 as well as the normal state properties.
  In case (i) the local pairs of "d"  electrons can move  
  via  a mechanism of virtual 
  excitations into empty c-electrons states. Such a mechanism  gives  rise 
  to the long range hopping of  pairs  of  "d"  electrons  
 (in analogy  to the  RKKY 
  interaction  for s-d mechanism in the magnetic equivalent).  
The superconducting properties are analogous to those of  a pure local 
pair superconductor \cite{{Micnas90},{NATO}}.
In  case  (iii),  on  the 
  contrary, we find a situation which is similar to the BCS case: pairs of 
  "c" electrons with opposite momenta and spins are exchanged via virtual 
  transitions into local pair states. 

The generic phase diagrams for different pairing  symmetries  
plotted as a function 
of the position of the LP level $\Delta_{0}$ at fixed $n$ 
are shown in Fig.2a-c.
In all the cases one observes a 
 drop in the superfluid stiffness (and in the  KT transition 
temperature) when the bosonic level reaches the bottom 
of the $c$-electron band and the system approaches the LP limit.  
In the opposite, BCS like limit, $T_{c}^{KT}$ approaches
asymptotically 
$T^{MFA}_{c}$,
 with  a narrow fluctuation regime.
Between the KT and MFA temperatures,  the {\it phase fluctuation effects} 
are important. 
In this regime a pseudogap in the c-electron spectrum will develop and 
the normal state of LP and itinerant fermions  
can exhibit non-Fermi liquid properties \cite{jrmr96}.

The  analysis of the Mixed-LP crossover indicates that  when the LP 
level is lowered and  reaches the bottom of the fermionic band 
an effective attraction between fermions becomes 
strong, since it varies as $I^2/(2\Delta_{0}-2\mu)$ and $\mu\approx \Delta_{0}$
(Eq. \ref{effint}). 
In this regime the density of $c$ electrons is low and 
 formation of bound
$c$-electron pairs occurs.  
It gives rise 
to an energy gap  in the single-electron spectrum {\it independently of the pairing
symmetry}.
 We  evaluated the binding energies of $c$- 
electron pairs and
found that $T_{c}^{MFA}$ scales with the  half of their binding energy for 
$\Delta_{0}\le 0$.
The superconducting transition temperature  is here always 
much lower than the $c$-pair formation temperature ($T_{c}^{MFA}$) and
rapidly decreases with $|\Delta_{0}/D|$. In such a case, the superconducting
state can be  formed by two types of coexisting (composite) bosons: 
preformed $c$- electron pairs and LP's \cite{unpublished,ohashi}.

Comparing $T_{c}$ {\it vs} $\Delta_{0}$ plots for various pairings   one finds that in
the case of nn hopping only, the $d$ and $s$ -wave pairings are favorable for
higher concentration of $c$-electrons, while the $s^{*}$-wave can be stable at
low $n_{c}$.  This is clearly seen from Fig.3, where the plots of $T_{c}^{MFA}$
{\it vs} $\Delta_{0}$  for $s, s^{*}$ and $d$-pairings are given for $n=1,
|I_{0}|/D=0.25$, and from Fig.4 which shows variation of $n_{c}$ and 
$n_{B}$  with $\Delta_{0}/D$ for the same values of $n$ and $\Delta_{0}/D$
for $d$ -wave pairing (for $s$ and $s^{*}$ -wave symmetries the plots of
$n_{c}$ and $n_{B}$  are very
similar).

A strongly  nonmonotonous behavior of $T_{c}$ {\it vs} $\Delta_{0}$ 
for $s^{*}$-wave pairing and $n>1$ (cf. Fig.2b) reflects the fact that this type
of pairing in a single band model is realized for low electron (or hole
densities), if $t_{2}=0$ \cite{Micnas90}.
In general, with increasing $n_{c}$ $(\Delta_{0}/D$) one can have a sequence of
transitions: $s^{*}\rightarrow d$ (or $s$)$\rightarrow s^{*}$ (if $1<n\leq2$),
or the transitions: $s^{*}\rightarrow d $ (or $s$) (if $n\leq1 $).\\
The nnn hopping $t_{2}$ (with opposite sign to $t$) can strongly enhance $T_{c}$
for $d$-wave symmetry, moreover it favors the $d$ and $s$-wave pairings for
lower values of $n_{c}$ \cite{unpublished}.

The region between $T_{c}^{MFA}$ and $T_{c}^{KT}$, where the system can exhibit a
pseudogap, expands with increasing intersubsystem coupling $|I_{0}|$. 
It is seen from Fig.5 that, except for $|I_{0}|/D\ll 1$ the coupling dependences
of $T_{c}^{MFA}$ and $T_{c}^{KT}$ are qualitatively different.
 $T_{c}^{MFA}$ is an increasing function of $|I_{0}|$ for all the pairing
 symmetries. On the other hand, $T_{c}^{KT}$ {\it vs} $|I_{0}|$ increases first, goes
 through a round maximum and then decreases (similarly as it is observed in the
 attractive Hubbard model\cite{Den}). The  position of the maximum 
 corresponds to  the intermediate
 values of $|I_{0}|/D$ and it depends on the pairing symmetry as well as the values
 of $\Delta_{0}/D$ and $n$. For large  $|I_{0}|$, the 
 $T_{c}^{KT}$ are close
 to the upper bound for the phase
 ordering temperature which is given by $\pi\rho_{s}(0)/2$.

As for the evolution of  superconducting properties with increasing $n$
one finds three possible types of density driven changeovers
\cite{unpublished,czart01}: 
(i)  for $2\geq \Delta _0/D\geq 0$, "BCS"$\longrightarrow $  
Mixed $\longrightarrow $ "BCS";
(ii) for $\Delta_{0}/D>2$: "BCS" $\longrightarrow$ "LP" and 
(iii) for $\Delta_{0}/D<0$: "LP" $\longrightarrow$ "BCS".
Only if the LP level is
deeply located below the bottom of  $c$-band,
the system remains in the LP regime for any $n\leq 2$.

 Some of our findings  can be qualitatively 
 related to experimental 
   results for the cuprate HTS where a pseudogap exists. It has
   been suggested  by ARPES experiments, that for underdoped cuprates 
   the Fermi surface in the 
   pseudogap phase is truncated around the corners 
   due to the formation of preformed (bosonic) pairs with charge $2e$, 
   whereas the "electrons" on the
   diagonals 
    remain  unpaired [7a].
   In the present two component model such a situation is obtained 
   when LP's and $c$-electrons coexist in the mixed regime. 
   The linear $T$-dependence of the 
   superfluid density has been observed 
   experimentally in  copper oxides and also in 
   several organic superconductors. 
   This points 
   to an order parameter 
   of $d_{x^2-y^2}$-wave symmetry and existence of nodal quasiparticles. 
   In the present model the gap ratio is nonuniversal and can  deviate strongly 
   from BCS predictions (particularly in the d-wave case 
  for which it is  always  enhanced) \cite{unpublished}.  
   This feature  is also found for
   several  exotic superconductors.   
   The Uemura plots i.e., the  $T_{c}$ 
{\it vs} zero-temperature phase stiffness 
$\rho_{s}(0)$ (and the scaling $T_{c}\sim \rho_{s}(0)$) 
 reported for  cuprates and organic
   superconductors 
   can be reproduced within the model for 
   extended $s$- and $d$-wave order parameter symmetry 
   (\cite{rmsrbt}, Figs.11,13 in Ref.\cite{unpublished}).\\
Finally, we point out that  to relate more quantitatively 
the present two-component scenario
  to the phase diagram of
 cuprates one has to consider doping dependences of
 the model parameters. 
\section*{Acknowledgements} 
This work was supported in part by the Polish State Committee for 
Scientific Research (KBN), Project No. 2 P03B 154  22.

\vfill

\newpage
\begin{figure}
\begin{center}
\includegraphics[scale=0.6]{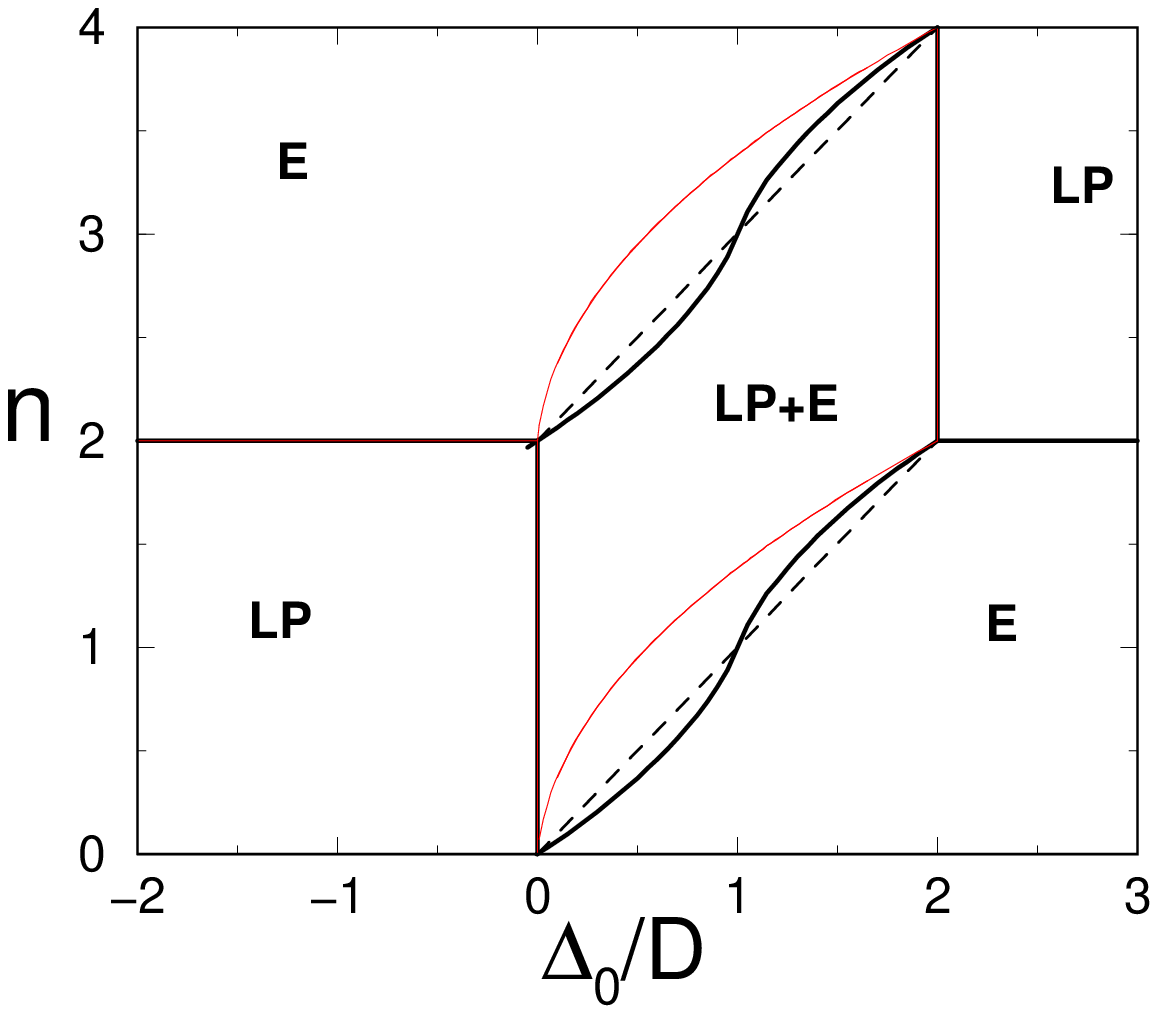}
\end{center}
\caption{\small {Ground state diagram of the model (1) as a function of $n=n_{c}
+2n_{B}$ versus $\Delta_{0}/D$ ($D=zt=4t$) for $|I_{0}|=0$,
plotted for a  square lattice with $t_{2}=0$ (thick solid lines), rectangular
DOS (dashed lines) and with 
 $t_{2}/t=-0.45$ (thin lines). 
 LP - nonmetallic state of LP's ($2n_{B}=n, n_{c}=0$ for $n<2$, 
 ~$2n_{B}=2-n, n_{c}=2$ for $n>2$), 
 E - metallic state of $c$ -electrons ($n_{c}=n, 2n_{B}=0$ for $n<2, 
 n_{c}=2-n, 2n_{B}=2$ for $n>2$), LP+E - mixed regime ($0<n_{c}, 2n_{B}<2$). }}
\end{figure}
\newpage 
 \begin{figure}
\includegraphics[scale=0.55] {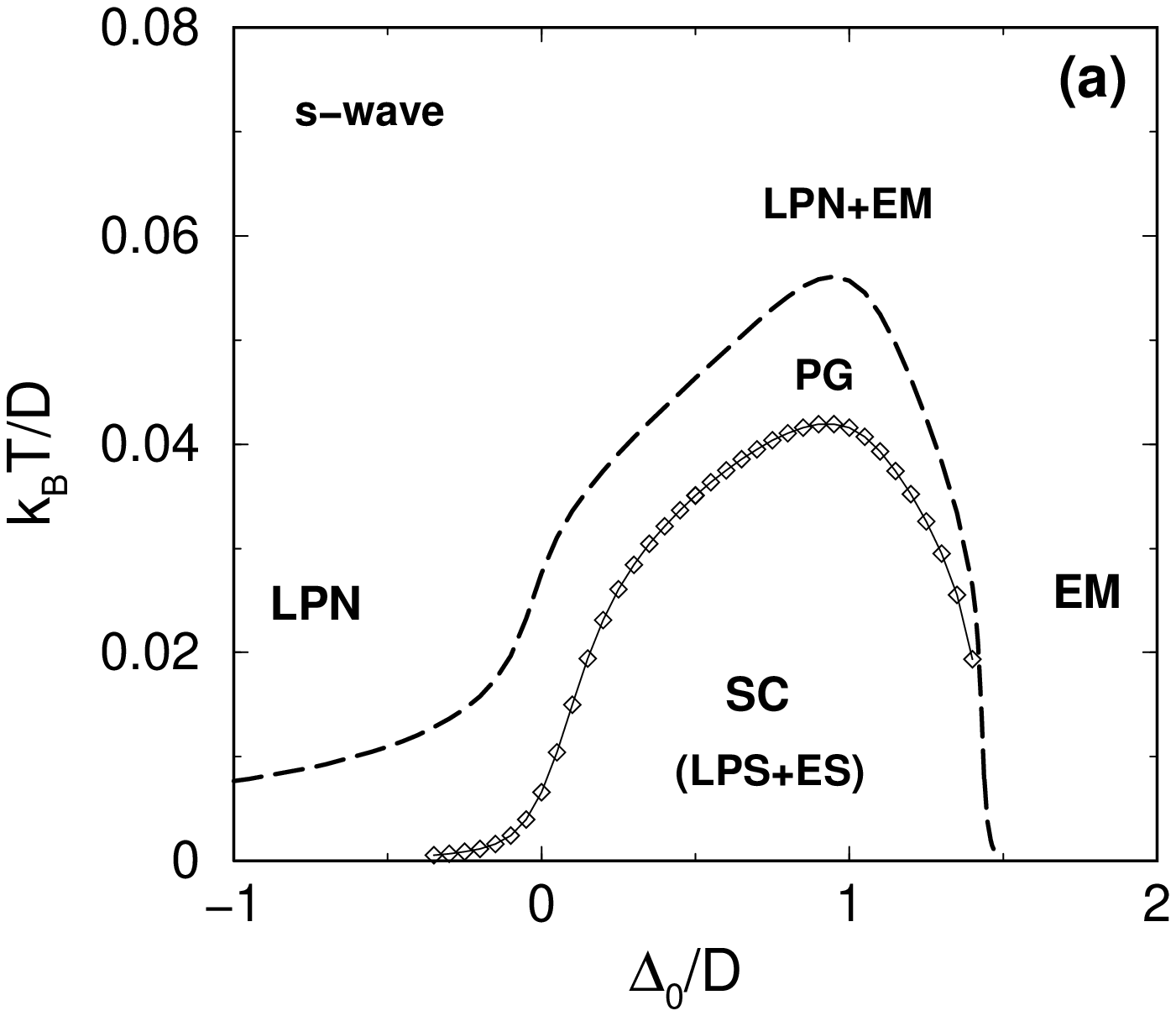}
\includegraphics[scale=0.55] {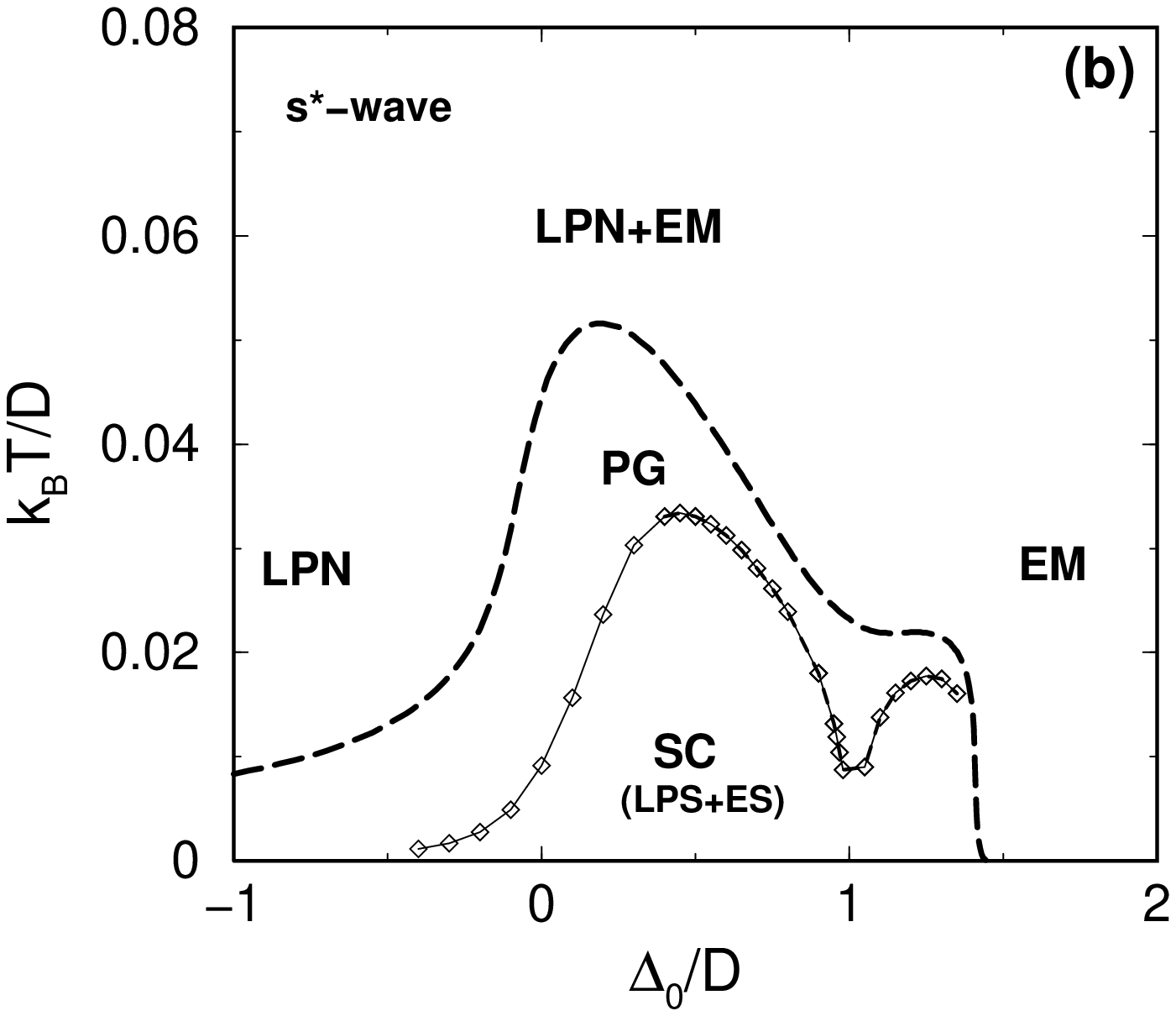}
\begin{center}
\includegraphics[scale=0.55] {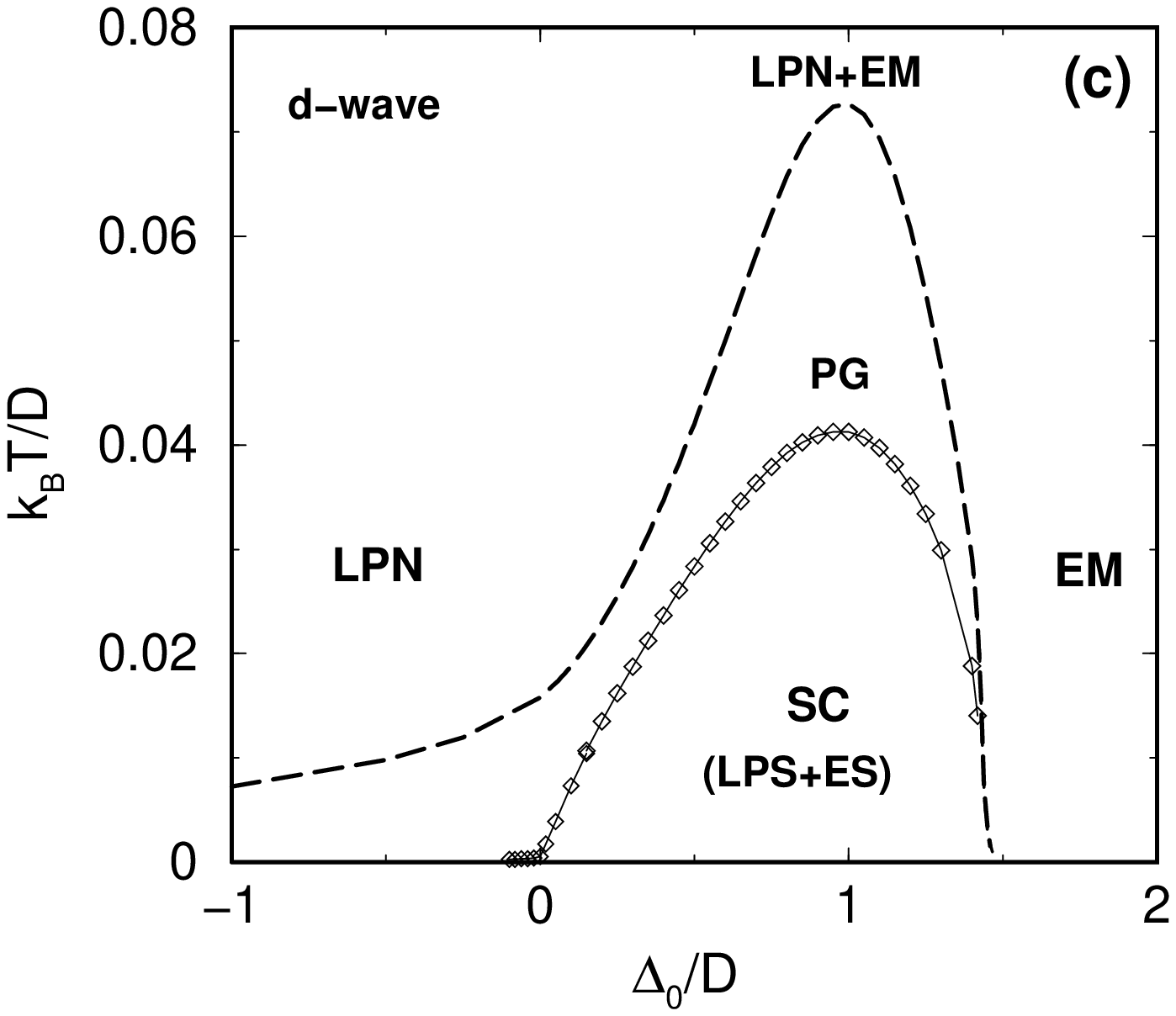}
\end{center}
\caption{{\small Phase diagrams of the induced pairing model as a function of
$\Delta_{0}/D$ derived for (a) $s$-wave, (b)- extended $s$-wave, 
(c) $d$-wave symmetry ($t_{2}=0)$.  
In all the figures the dashed line 
shows the BCS-MFA transition
temperature, while the line with diamonds the KT transition temperature
calculated for $n=1.5, |I_{0}|/D=0.25$. 
LPN--normal state of predominantly  LP's, EM--electronic metal, 
LPS+ES--superconducting (SC) state, PG -- pseudogap region. }}
\end{figure}
\begin{figure}
\begin{center}
\includegraphics[scale=0.6]{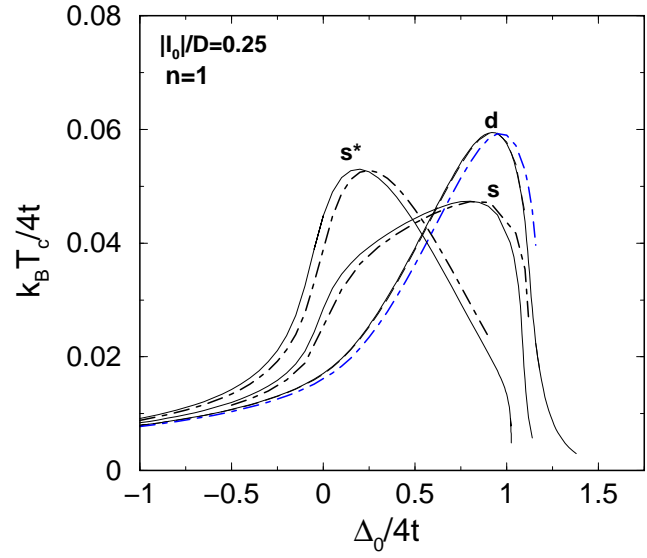}
\end{center}
\caption{{\small $T_{c}^{MFA}$ as a function $\Delta_{0}/D$ 
for different pairing symmetries plotted for $n=1, |I_{0}|/D=0.25, t_{2}=0$.
Solid lines are the results for the $d=2$ square lattice, while the dashed ones
- for the quasi 2D case with $t_{\perp}/t=0.1$}}
\end{figure}
\begin{figure}
\begin{center}
\includegraphics[scale=0.55]{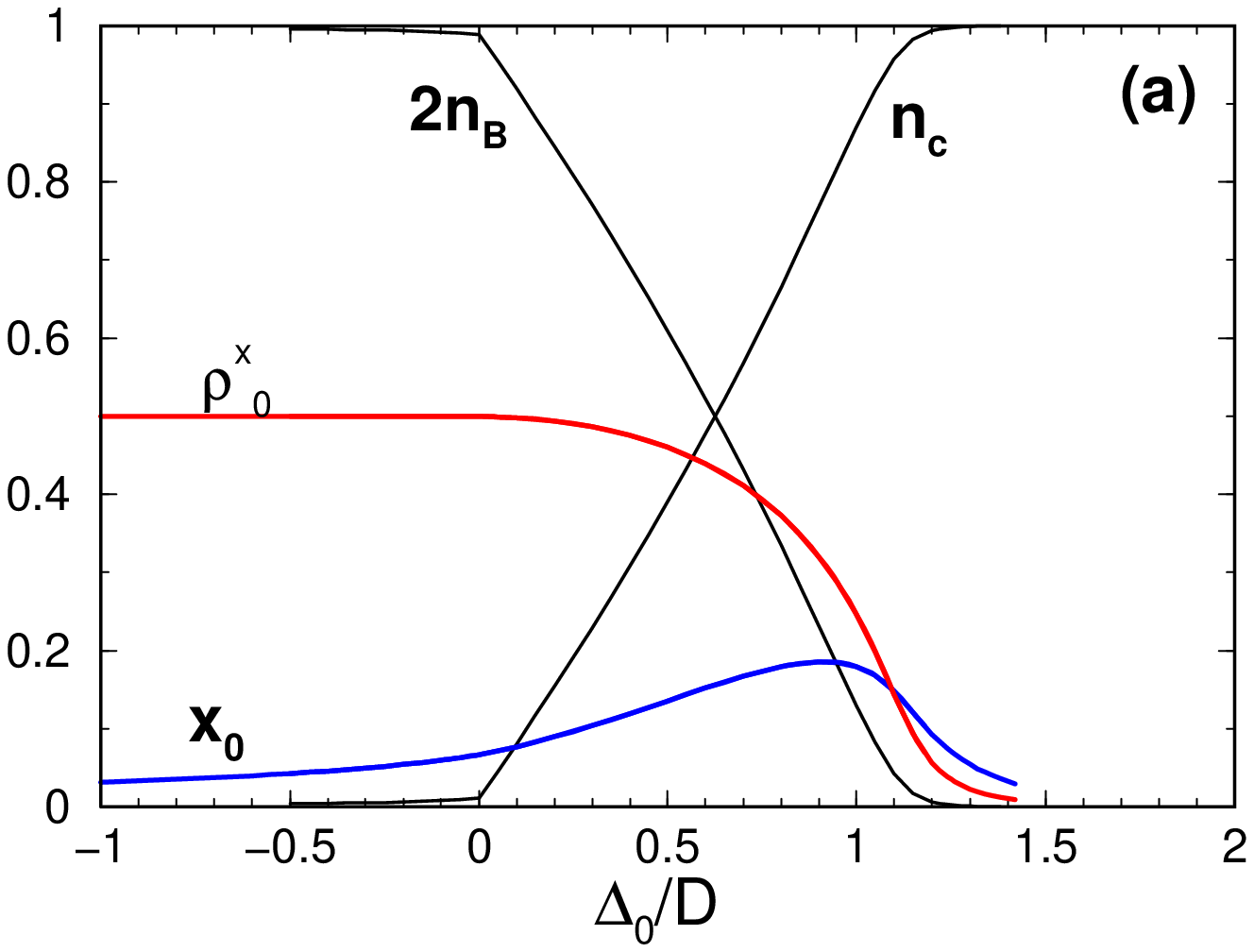}
\end{center}
\begin{center}
\includegraphics[scale=0.55]{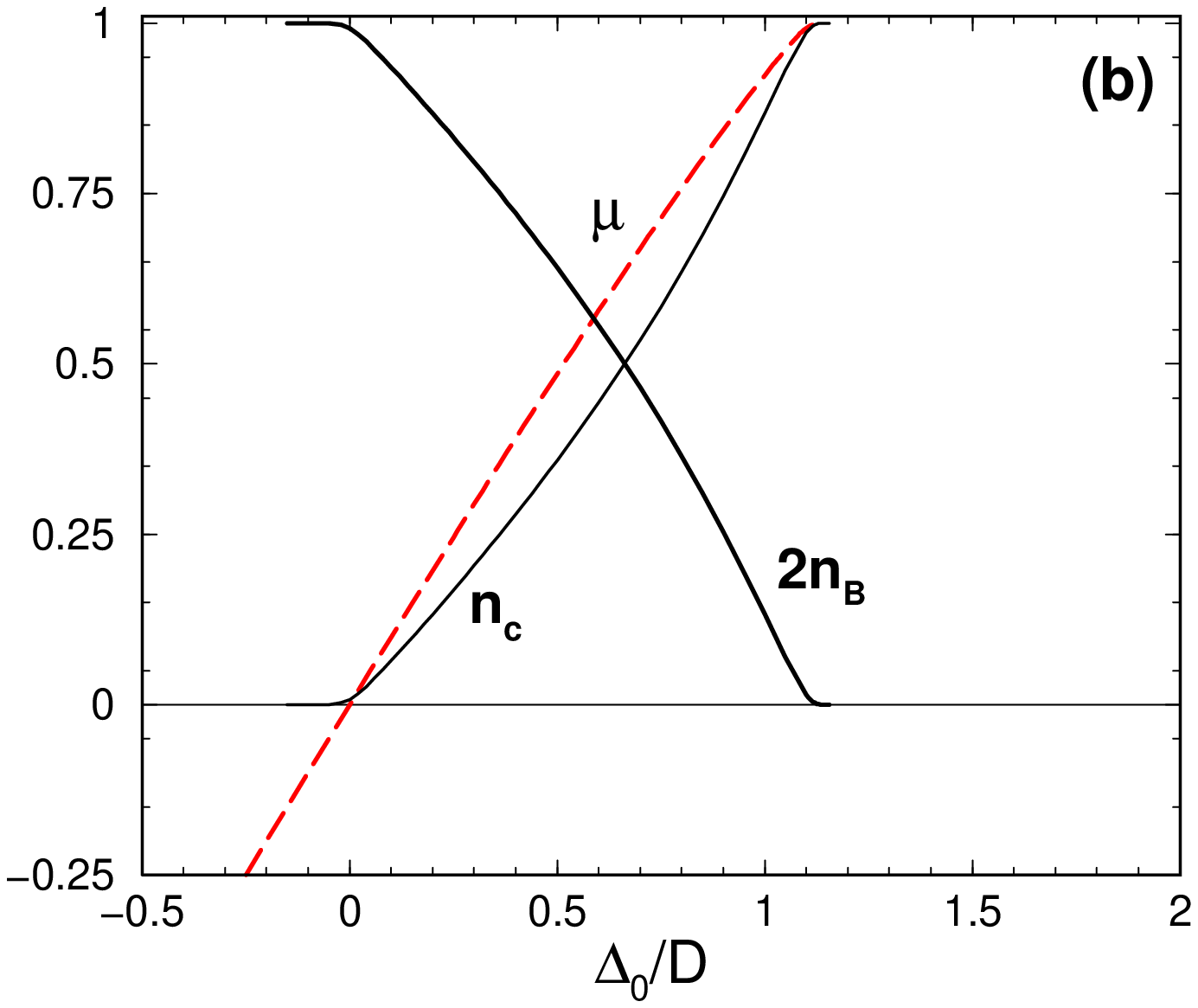}
\end{center}
\caption{{\small (a) Variation of $n_{c}$ (concentration of $c$-electrons), $n_{B}$
(concentration of LP's) and  superconducting order parameters $\rho_{0}^{x},
x_{0}$ at $T=0$ as  a function of $\Delta_{0}/D$, for $n=1, |I_{0}|/D=0.25, 
t_{2}=0$. $d$-wave pairing.
(b) The corresponding plots of $n_{c}, n_{B}$ and  $\mu/D$ at
$T_{c}^{MFA}$.}}
\end{figure}
\begin{figure}
\begin{center}
\includegraphics[scale=0.6] {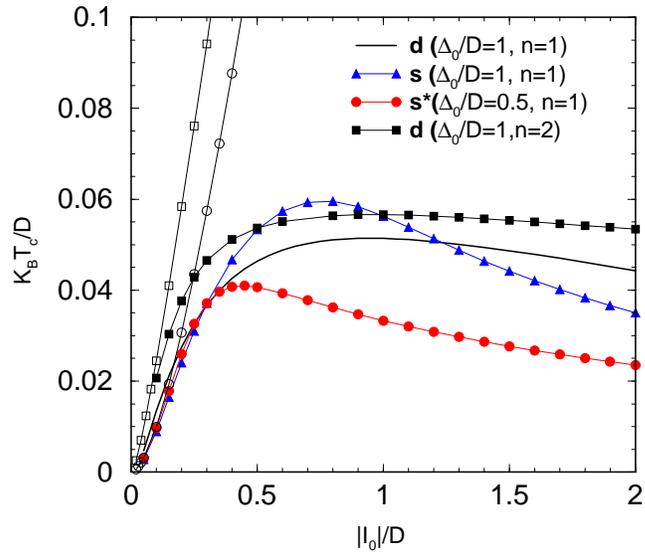}
\end{center}
\caption{{\small Temperatures $T_{c}^{MFA}$ and $T_{c}^{KT}$ 
as a function of $|I_{0}|/D$ for various
types of pairing for a 2D lattice ($t_{2}=0)$. The lines with full symbols and thick solid
line denote the  KT
transition temperatures. The line with empty squares is  $T_{c}^{MFA}$ for
$d$-wave pairing ($\Delta_{0}/D=1, n=2$), while the line with open circles is
$T_{c}^{MFA}$ for $s^{*}$-wave pairing ($\Delta_{0}/D=0.5, n=1$). }}
\end{figure}

\end{document}